\documentstyle[prl,aps]{revtex}
\begin{document}
\draft
\input epsf
\twocolumn[\hsize\textwidth\columnwidth\hsize\csname
@twocolumnfalse\endcsname
%
%
\title{{\hfill \small Desy-01-038, DSF-11-2001, quant-ph/0103144}\\ $~$\\
         When does a detector click?}

\author{R. Brunetti${}^1$ and K. Fredenhagen${}^2$}
\address{$~$\\${}^1$ Dip. di Scienze Fisiche, Univ. Napoli 
''Federico II", Com. Univ. Monte Sant'Angelo, Via Cintia, I-80126 Napoli, 
Italy.}
\address{${}^2$ II Inst. f. Theoretische Physik, Universit\"at Hamburg, 149 
Luruper Chaussee, D-22761 Hamburg, Germany.}

\maketitle

\begin{abstract}
{We propose a general construction of an observable 
measuring the time 
of occurence of an effect in quantum theory. Time delay in
potential scattering theory is computed as a 
straightforward application.}
\end{abstract}

\vskip2pc]

\section{Introduction}
Time measurements play an important role in experiments, but in
quantum theory the corresponding observables are not easy to find. The
difficulties are connected with the fact that the dual observable, the
energy, has a nontrivial spectrum, on which shifts are in general not
well defined.

It is well known and easy to see that no selfadjoint operator exists
which decribes the measurement of time \cite{Pauli}. 
Therefore one has to rely on a
more general concept of observables, and a natural option is the 
concept of positive operator valued measures \cite{Neumark}. 
This concept has meanwhile often been applied to the problem of time 
measurements (see e.g. \cite{Bush}). What is missing, however, so far we know, 
is a discussion of the {\sl effect} whose occurence time is described. 
A description of the latter is the main contribution of this letter.
%

\section{The Construction of a Time Operator}
Let $A>0$ be a bounded operator on a Hilbert space $\mathcal H$ 
which measures the occurence of an
effect at time $t=0$. In Quantum Mechanics one might think of the
projection operator $P(x\in M)$ for the position being in the region
$M$, see e.g. \cite{AJM}, in Quantum Field Theory a choice would be an 
Araki-Haag counter \cite{ArakiHaag}, 
i.e. an almost local operator with vanishing vacuum expectation value, an 
(unbounded) example, in free massive field theory, being the partial 
number operator $a(f)^*a(f)$ where $a(f)$ denotes the 
annihilation operator for a particle with a smooth momentum 
space wave function $f$. 

Given a state $\omega$ and a time evolution $\alpha_t$, one may
consider the expectation values $\omega(\alpha_t(A))$ as a nonnegative 
function of $t$ (assumed to be continuous). At times $t$ when this
function is big the occurence of the effect will be more probable 
than at times $t$
when it is small, and one may ask whether one can derive from this a
probability distribution. It may happen that 
$\int\omega(\alpha_t(A)) dt =\infty$, so the time spent in the
detector is infinitely long, and a probability distribution cannot be
defined (in this case an arrival time might possibly be introduced
(see e.g. \cite{Werner})). It may also happen that $\omega(\alpha_t(A))=0$ for
all $t$, which means that the effect never takes place; also then it
does not make sense to discuss a probability distribution. But in the
intermediate case when the integral is finite a probability
distribution can be found.

The question arises whether such a probability distribution can be 
written as the expectation value of a positive operator valued measure. 
This means essentially that the normalization should be done on the level 
of operators, not on the level of expectation values.

We start from the integrals
$$
B(I)=\int_I\alpha_t(A) dt \ ,
$$
where $I$ is some bounded interval of the real line $\mathbf R$. 
Hence $B(I)$ is a positive bounded operator, and
$B(J)\ge B(I)$ if $J\supset I$. We want to construct the 
operator
$$B=\int_{\mathbf R}\alpha_{t}(A)dt\ .$$
Up to the normalization of $A$, $B$ may be 
interpreted as the total time duration of the effect.

For this purpose we consider the operators $(B(I)+1)^{-1}$. 
They form a decreasing net of positive bounded operators with greatest
lower bound $C\le 1$.
Let $\mathcal{H}_{\infty}$
be the kernel of $C$ and ${\mathcal H}_{0}$ the joint kernel of 
$\alpha_t(A)$. In the corresponding states the time of occurence of the effect 
is infinite, resp. $0$. Let $\mathcal{H}_{\text{finite}}$ 
be the orthogonal complement of 
${\mathcal H}_{\infty}+{\mathcal H}_{0}$ in $\mathcal H$.
On $\mathcal{H}_{\text{finite}}$ we define $B$ as the operator 
$$
 B=C^{-1}-1\ .
$$
$B$ is a positive selfadjoint, in general unbounded operator, which dominates 
all operators $B(I)$ (restricted to operators in $\mathcal{H}_{\text{finite}}$). 
We set
$$
P(I)=B^{-\frac{1}{2}}B(I)B^{-\frac{1}{2}}\ ,
$$
and get positive operators bounded by 1 with
$$
P(I\cup J)=P(I)+P(J)\ ,
$$
for disjoint intervals $I$ and $J$ and with
$$
P({\mathbf R})=1\ .
$$
One may readily check that the countable additivity property of $P$ is also 
fulfilled. So we obtained a positive operator valued measure 
which transforms in the right way under time translations,
$$
\alpha_t(P(I))=P(I+t)\ .
$$

To compare our concept with other definitions of time measurements we 
consider the first moment of the measure (the time operator 
associated to $A$),
$$
T_{A}=\int t P(dt)\ .
$$
We assume that the time translations are implemented by 
unitaries $e^{itH}$ with a selfadjoint Hamiltonian $H$
with absolutely continuous spectrum and that the Hilbert space 
of our model can be represented as the space of sections of a 
(trivial) bundle of finite dimensional Hilbert spaces over the spectrum of $H$. 

Let $A$ be a positive operator with a smooth integral kernel $a(E,E')$ 
with $a(E,E)>0, \forall E$. Then the integral kernel of $B$ is
\begin{eqnarray*}
b(E,E')&=&\int e^{it(E-E')}a(E,E') dt \\
&=& 2\pi \delta(E-E')a(E,E')\ .
\end{eqnarray*}
For $P(I)$ we find the integral kernel
$$
\frac{1}{2\pi}c(E,E')\int_I e^{it(E-E')} dt\ ,
$$
with 
$$
c(E,E')=a(E,E)^{-\frac{1}{2}}a(E,E')a(E',E')^{-\frac{1}{2}}\ .
$$
For the first moment (the time operator) we obtain on smooth sections 
$\Phi$ with compact support 
which vanish at the boundaries of the spectrum of $H$
$$
T_{A}\Phi(E)=(-i\partial_E +d_{A}(E))\Phi(E)\ ,
$$
with
$$
d_{A}(E)=-i\partial_{E'}c(E,E')|_{E'=E}\ .
$$
Note that $d_{A}$ is hermitian because of the positivity of 
$c$.

We see that the time operator $T_{A}$ has the expected form as a 
covariant derivative in our bundle. Different effects give rise 
to different connections $d_{A}$.

From this result we can immediately obtain the observable which 
measures the time in between two effects. If $d_1$ and $d_2$ are 
the corresponding connections (we may assume that they 
are bounded, so the domain of the time operators in question coincide), 
the transition time between both effects is
$$
d_1-d_2\ .
$$
The transition time is therefore (under the above conditions) a 
selfadjoint bounded operator showing none of the problems associated 
with the time operator.
(This is in conflict with the statement
in \cite{Unruh} that also transition times have an intrinsic 
uncertainty. The reason for this discrepancy, as far we can 
see, is that in \cite{Unruh} the transition time is 
analyzed only in eigenstates of the momentum whereas the 
eigenstates of the transition time as defined above are superpositions of 
different momentum eigenstates with the same energy.) 
\section{Application to Scattering Theory}
A famous example for an analysis of times in quantum theory
is the time 
delay in scattering theory (see, e.g., \cite{AJM}\cite{Martin}). 
Here a general formula can be derived which is valid for all 
interactions. In the literature
\cite{Martin} it is done using the so-called sojourn time. 
The sojourn time is defined as the expectation value of the 
operator $B$ introduced above where the effect $A$ is the 
projection operator characterizing the probability that the 
particle is in a certain region of space which eventually 
tends to the whole space. One then compares the sojourn times of 
the interacting case with the noninteracting case. 
This analysis requires  delicate arguments of convergence. 
We want to show that our concept leads to a much more 
direct and elementary derivation.

Let us consider a typical situation 
of scattering in quantum mechanics: The initial state $\phi_{\text{in}}$ 
evolves according to the interacting Hamiltonian and asymptotically 
tends to $\phi_{\text{out}}= S \phi_{\text{in}}$ where $S$ 
is the scattering matrix (assumed to be unitary). The
time delay of the event measured by the positive operator $A$ is given by 
the formula
$$
\langle \phi_{\text{out}},T_A\phi_{\text{out}}\rangle -
\langle \phi_{\text{in}}, T_A \phi_{\text{in}}\rangle  \,
$$
(provided the domain of $T_{A}$ is invariant under $S$). 
Hence,
\begin{eqnarray*}
t_{\text{delay}}& =& \langle \phi_{\text{in}}, (S^{-1}T_A S-T_A)
\phi_{\text{in}}\rangle \\
&=& \langle \phi_{\text{in}}, S^{-1}[T_A,S]\phi_{\text{in}}\rangle  \ .
\end{eqnarray*}

According to what has been exposed in the previous section we 
obtain 
\begin{eqnarray*}
t_{\text{delay}}(E)&=& S(E)^{-1}(-i\partial_E)S(E)\\ 
&+& 
S(E)^{-1}[d_{A}(E),S(E)]\ ,
\end{eqnarray*}
so the time delay is just the change of the connection 
under the action of the on-shell scattering matrix (considered as a gauge 
transformation). If $d_{A}$ commutes with $S$ (this is 
always the case when the energy spectrum is nondegenerate) 
one obtains the well known formula of Eisenbud and Wigner 
(see \cite{Wigner}, but especially \cite{Martin} and \cite{AJM}).

As a simple explicit example let us 
analyze the case of a central repulsive and short range potential. 

The radial wave function of a scattering state of the particle  
with sharp angular momentum has the form
$$
\psi(t,r)=\int e^{-itE}u_k(r)\varphi(k) dk\ , \ E=\frac{k^2}{2m}\ ,
$$
where the functions $u_k$ are solutions of the radial 
Schr\"odinger equation with the asymptotic behavior for $r\to \infty$
$$
u_k(r)\sim e^{-ikr} + e^{ikr+i\delta(k)}\ ,
$$
with a real function $\delta$ (the phase shift).

We first have to find an observable which indicates the passage of the 
particle through a spherical shell, of thickness $\rho$ (which 
eventually will tend to zero) with a 
sufficiently large radius $R$, in outward direction. 
Such an observable is 
$$
A=Q^*PQ \ ,
$$
where $P$ is the projection onto states with positions inside the 
shell and where $Q$ simulates the (nonexistent) projection on 
states with positive radial momentum. We may choose
$$
Q=-i\partial_r (2mH)^{-\frac{1}{2}}-1 \ ,
$$
where $\partial_r$ is the derivative operator with boundary 
condition $\psi(r=0)=0$ (then $-i\partial_r$ is maximally 
symmetric but not selfadjoint). 
Note that $Q$ is bounded and that it selects for large $r$ 
the outgoing component of $u_k$.

We now insert $A$ into the formulas derived before and get 
$$
a(E,E')=\frac{m}{\sqrt{kk'}}\langle Qu_k,PQu_{k'}\rangle\ ,
$$
\begin{eqnarray*}
&c(E,E')&=\\
&&\langle Qu_k,PQu_k\rangle^{-\frac{1}{2}}
        \langle Qu_k,PQu_{k'}\rangle
        \langle Qu_{k'},PQu_{k'}\rangle^{-\frac{1}{2}}\ ,
\end{eqnarray*} 
\begin{eqnarray*}
d(E)&=-\frac{im}{2k}\langle Qu_k,PQu_k\rangle^{-1}
                  &\bigl(\langle Q\partial_ku_k,PQu_k\rangle \\
&& -\langle Qu_k,PQ\partial_k u_k\rangle\bigr)\ .
\end{eqnarray*}
Provided $r$ is sufficiently large, we can replace $u_k$ 
by its asymptotic form and find
$$
c(E,E')=\frac{\sin \rho(k-k')/2}{\rho(k-k')/2}
e^{iR(k-k')}e^{-i(\delta(k)-\delta(k'))}\ .
$$
We see that we can safely take the limit $\rho\to 0$.

As a test we evaluate our positive operator valued measure in a state
$$
\psi(r)=\int \varphi(k)u_k(r) dk\ ,
$$
where $\varphi$ has support in a small neighbourhood.
The probability density of passage times is given by the integral
$$
p(t)=\int\overline{\varphi(k)}\varphi(k')
e^{i((k-k')R+\delta(k)-\delta(k')+t(E-E'))}\ dk\ dk'\ .
$$
Since the support of $\varphi$ is small, we may linearize $\delta$ 
as a function of $E$ and find that 
$p(t+\partial_E \delta)$ is independent of the interaction. 
Hence (in this approximation) the distribution of the times 
when the particle passes through the shell in outward 
direction is delayed and we get the well known time delay
$$
t_{\text{delay}}=\partial_E \delta\ .
$$
The same conclusion can also be obtained from the general definition
of the time delay with the time operator $T_A$ as derived previously. 
A straightforward computation gives
$$
d_{A}(E)=\frac{mR}{k}-\partial_E \delta\ ,
$$
which yields again the logarithmic derivative of the on-shell scattering 
matrix. Note, however, that the first moment of a positive operator valued 
measure does in general not contain all informations on the measure 
(in contrast to projection valued measures).

As a byproduct we see that the connection $d_{A}$, in the 
case of the free time evolution, is a function of the energy 
(independent of the angular momentum). Therefore it 
commutes with the scattering matrix also for noncentral potentials, 
and we obtain the Eisenbud and Wigner formula in the general case.

\section{Conclusions and outlook}
We presented a general construction of a
time operator measuring the occurence time of an effect and used it for a new 
derivation of the well known formula for time delay in scattering theory.
Our approach may be generalized in several directions. 
The first is the study of coincidence arrangements of detectors, 
either in quantum mechanics or in quantum field theory, leading to a 
distribution of multiple times resembling the general framework of 
the consistent histories approach. Another generalization is an analysis  
of times in a periodic or quasiperiodic situation related
to bound states of the Hamiltonian. There the positive operator valued 
measure is concentrated on a compact space on which time translations 
act, quite similar to Bellissard's action of translations on 
homogenous disordered systems \cite{Bellissard}. 
One may also use a similar construction to 
characterize the localization of an event in {\sl spacetime}. 
This may be compared with the somewhat different 
ansatz in \cite{Giannitrapani,Toller}. We hope to 
report soon elsewhere on these interesting topics.

\acknowledgements
The authors acknowledge the support of the EU, INFN-sez. Napoli, DFG and 
MURST, whose grants allowed the collaboration during the last years.
 

\end{document}